\documentclass[prd,twocolumn,showpacs,superscriptaddress,groupedaddress,nofootinbib]{aastex62}

\usepackage{graphicx}
\usepackage{amsmath,amssymb} 
\usepackage{color}

\usepackage{xspace}

\newcommand{\PlotPath}{./}

\begin{document}


\title{Observational implications of lowering the LIGO-Virgo alert threshold}
\author{Ryan Lynch}\email{ryan.lynch@ligo.org}
\affiliation{Massachusetts Institute of Technology, Cambridge, MA, 02139, USA}

\author{Michael Coughlin}
\affiliation{Division of Physics, Math, and Astronomy, California Institute of Technology, Pasadena, CA 91125, USA}

\author{Salvatore Vitale}
\affiliation{Massachusetts Institute of Technology, Cambridge, MA, 02139, USA}

\author{Christopher W. Stubbs}
\affiliation{Department of Physics, Harvard University, Cambridge, MA 02138, USA}
\affiliation{Department of Astronomy, Harvard University, Cambridge MA 02138, USA}

\author{Erik Katsavounidis}
\affiliation{Massachusetts Institute of Technology, Cambridge, MA, 02139, USA}


\begin{abstract}

The recent detection of the binary-neutron-star merger associated with GW170817 by both LIGO-Virgo and the network of electromagnetic-spectrum observing facilities around the world has made the multi-messenger detection of gravitational-wave events a reality.
These joint detections allow us to probe gravitational-wave sources in greater detail and provide us with the possibility of confidently establishing events that would not have been detected in gravitational-wave data alone.
In this paper, we explore the prospects of using the electromagnetic follow-up of low-significance gravitational-wave event candidates to increase the sample of confident detections with electromagnetic counterparts.
We find that the gravitational-wave alert threshold change that would roughly double the number of detectable astrophysical events would increase the false-alarm rate by more than 5 orders of magnitude from 1 per 100 years to more than 1000 per year.
We find that the localization costs of following-up low-significance candidates are marginal, as the same changes to false-alarm rate only increase distance/area localizations by less than a factor of 2 and increase volume localization by less than a factor of 4.
We argue that EM follow-up thresholds for low-significance candidates should be set on the basis of alert purity ($P_\text{astro}$) and not false-alarm rate.  
Ideally, such estimates of $P_\text{astro}$ would be provided by LIGO-Virgo, but in their absence we provide estimates of the average purity of the gravitational-wave candidate alerts issued by LIGO-Virgo as a function of false-alarm rate for various LIGO-Virgo observing epochs.

\end{abstract}

\section{Introduction}\label{Sec.Intro}

The August 2017 detection of GW170817 was an event of many firsts.
Not only was it the first binary-neutron-star merger detected~\citep{GW170817} by the LIGO-Virgo detector network~\citep{StandardAdvancedLIGO,StandardAdvancedVirgo,DistanceCalc2013}, it was the first gravitational-wave (GW) event confidently detected by both ground-based GW detectors and electromagnetic (EM) observatories~\citep{EMFollow170817}.
While the detection of GW170817 could be confidently established by GW-detector data alone, the joint EM detection enabled a vast array of rich physical insights, such as the association of short gamma-ray bursts with binary-neutron-star mergers~\citep{GRB170817}, a new procedure for constraining the value of the Hubble parameter $H_0$~\citep{HubbleConstant}, and evidence of heavy-element nucleosynthesis~\citep{Swope,DECAM,DLT40,LasCumbres,VISTA,MASTER}.
The high signal-to-noise ratio of GW170817 and the clarity with which it could be distinguished in both GW and EM data aided the wealth of scientific information extracted by studying it.
However, given an astrophysical (uniform-in-volume) population of such sources, we expect that quieter GW candidates might also make a non-negligible scientific contribution to both the GW and EM communities.
For example, if we were able to double the number of joint GW-EM detections by performing searches for low-significance candidates, we could decrease the uncertainty in GW-based measurements of $H_0$ by up to a factor of $\frac{1}{\sqrt{2}}$ (although the actual improvement may be lower on account of poor distance localization)~\citep{ChenH0}.

In this paper, we examine the extent to which searches for low-significance GW transients can augment the total ensemble of GW detections.
For the purpose of establishing a baseline, let us assume that the minimum false-alarm rate (FAR) at which GWs can be confidently detected by LIGO-Virgo alone is 1 per 100 years.
This FAR corresponds to the nominal LIGO-Virgo alert threshold proposed for issuing open public alerts in the third Advanced LIGO-Virgo observing run~\citep{O3Alerts}.
In effect, we will define any GW event with a FAR of greater than 1 per 100 years to be a low-significance event.
Under this assumption, we cannot claim low-significance LIGO-Virgo events as confident detections unless they are jointly detected by EM observations at a convincing significance.
In a sense, this method is the complement to the scenario where GW events with extremely small localization volumes enable the discovery of faint EM counterparts~\citep{ChenFindingTheOne}.

However, there are potential opportunity costs that EM observers must weigh when considering how many GW candidates they can reasonably follow-up.
By definition, following-up GW candidates at a higher FAR threshold means a greater number of false-alarm contaminations.
Furthermore, low-significance candidates are inherently faint in GW detectors, implying that they may not be well-localized for EM observations.
The combination of these two factors along with finite observational resources suggests that each EM follow-up campaign should determine its own GW-alert-purity threshold that optimizes its scientific goals.
In the remainder of this paper, we quantify the ingredients necessary for calculating these purity thresholds.
This analysis is complementary to work optimizing EM follow-up using tiling, time allocation, and scheduling methods, e.g., \citep{GhBl2016,CoSt2016a,SoCo2017,ChHu2017,RaSi2017}.

\section{Source and background rates in GW detectors}\label{Sec.SignalRates}

\begin{figure}[h!]
    \scalebox{.85}{\includegraphics[width=0.5\textwidth]{\PlotPath/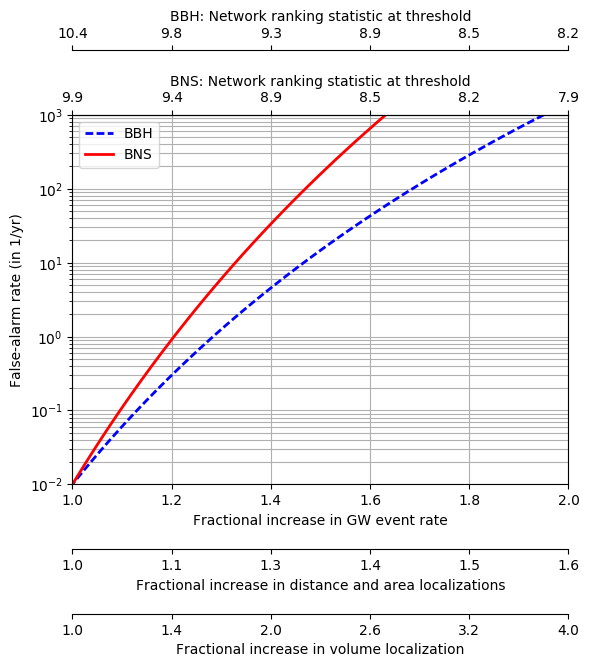}}
    \caption{The relationship of several statistics to the gravitational-wave (GW) search false-alarm rate (FAR).
    The top axes show the network ranking statistic ($\rho$) for binary-black-hole (BBH) and binary-neutron-star (BNS) searches, fit to the FAR versus $\rho$ curves presented in~\citep{PyCBCBack} using Eq.~\ref{Eq.Background}.
    A modest $\Delta\rho$ = 2 in the network ranking statistic increases the FAR by more than 5 orders of magnitude. 
    The bottom axes show the fractional increase (as compared to the value at a FAR of 1 per 100 years) of several quantities that scale as power laws in $\rho$: the GW event rate ($\propto \rho^{-3}$), the angular area and distance localizations (both $\propto \rho^{-2}$), and the volume localization ($\propto \rho^{-6}$).
    Increasing the FAR by 5 orders of magnitude from 1 per 100 years to 1000 per year increases the GW event rate/area localization/distance localization by less than a factor of two and increases the volume localization by less than a factor of 4.
    }
    \label{Fig.AllPowerLaws}
\end{figure}

All search algorithms for transient GW events follow the same basic hypothesis:  the signatures of GW events in every GW detector should be morphologically identical (once projection effects are taken into account) and time coincident, while detector noise need not be.
The noise in each GW detector is a superposition of a Gaussian bulk and non-Gaussian noise transients.
With low probability, this noise can mimic the appearance of GW events, which forms a background for the various search algorithms.
One such search algorithm is PyCBC~\citep{PyCBCAlgo,PyCBCCode}, which uses a bank of compact-binary-coalescence templates to rank GW detection candidates according to a network ranking statistic, $\rho$, that combines the candidate signal-to-noise ratio (SNR) with signal consistency tests~\citep{AllenChiSq,S6CBCAlgo}.
The main results of this paper are all extensions of the following points:  1.) background event rates fall off exponentially as a function of $\rho$, while 2.) GW event rates fall off less-steeply as a power law.  

The bulk of the background distribution of searches for GW transients falls off steeply as a function of $\rho$, meaning the FAR changes by orders-of-magnitudes over narrow ranges of $\rho$.
To quantify this more precisely, we explore the background for the two LIGO detectors:  one in Hanford, Washington, USA (H) and the other in Livingston, Louisiana, USA (L). 
The HL background rate (i.e., the FAR) decays roughly exponentially as a function of $\rho$ for FARs between 1 per 100 years and 1000 per year.
Thus, we can model the FAR as
\begin{equation}\label{Eq.Background}
	\text{FAR} = \text{FAR}_8 \times \exp\left[-\frac{\left( \rho - 8 \right)}{\alpha}\right]
\end{equation}
where $\text{FAR}_8$ is the FAR at $\rho = 8$, and $\alpha$ is scale-parameter that determines the steepness of the falloff.
This behavior is observed both for searches for compact-binaries~\citep{S6CBC,PyCBCBack} and searches for short-duration unmodeled GW events~\citep{O1AllSky}.
The steepness of this exponential falloff is determined by how easy it is for background events to mimic GWs in a given search.
Thus, searches for binary-neutron-star (BNS) events have a steeper falloff (smaller $\alpha$) than searches for binary-black-hole (BBH) events because a known time-frequency evolution is observed over longer durations for BNS events than for BBH events.
Likewise, searches for short-duration unmodeled GW events have less-steep exponential falloff than for either BNS or BBH events because the former's time-frequency evolution is inherently unknown and thus less constrained.

For this paper, we focus only on searches for BBH and BNS events~\citep{GW150914Pipelines,O1BBH,GW170817}, since both of these source-types have already been detected by LIGO-Virgo.
These detections have allowed for the sources' rates to be observationally established~\citep{GW170104,O1BBH,GW150914Rate,GW150914RateSupp,GW170817}.
As mentioned, BNS events have already been jointly detected by LIGO-Virgo and EM observers~\citep{EMFollow170817}, making them the most anticipated targets for low-significance efforts.
We do not present the results for short-duration unmodeled events since we do not have any direct measurements of their rate.
However, the relative results, obtained by normalizing out these unknown rates, are of similar magnitude to those for both BNS and BBH sources, resembling the results for BBH sources more closely.

We perform the exponential fit using the FAR versus $\rho$ relationship reported in \citep{PyCBCBack}, which represents the PyCBC search background for the HL data during the first Advanced LIGO-Virgo observing run. 
The results of this fit are shown in Fig.~\ref{Fig.AllPowerLaws}.
For BBH we find $\alpha = 0.18$ and $\text{FAR}_8 = 5500 \ \text{yr}^{-1}$, while for BNS we find $\alpha = 0.13$ and $\text{FAR}_8 = 30000 \ \text{yr}^{-1}$.
We assume that the slopes of these fits are representative of the BBH and BNS HL searches in current and future observing runs.
This assumption is based upon empirical results:  we observe similar fits in published results for both Advanced \citep{PyCBCBack} and initial \citep{S6CBC} LIGO-Virgo observing runs.
As we will soon see, the potential impact of searches for low-significance GW events is reduced as the slope of the searches' background distributions becomes steeper.
Thus, any improvements that make the GW searches' backgrounds less-heavily tailed (and hence steeper), such as adding a third detector like Virgo to potentially reduce the coincidence rate of high-$\rho$ noise transients, may further reduce the case for low-significance GW science.

We must likewise find a model to describe the rate of GW events versus $\rho$.
Assuming that GW events are distributed uniformly in volume, the {\it cumulative} rate of GW events exceeding an SNR threshold should roughly scale as SNR$^{-3}$ for Advanced-era GW detectors probing the low-redshift universe \citep{Schutz,UniversalSNR,SalvoSchutz} (although this scaling relation will break down for third-generation GW detectors that probe higher redshifts~\citep{SalvoSchutz}).
By construction, we expect that the network ranking statistic $\rho \sim \text{SNR}$ for real GW events~\citep{PyCBCBack}.
As the Advanced GW detectors improve in sensitivity, the overall rate of GW events being observed will increase accordingly.
We can estimate the rate of GW events for a given observing epoch as
\begin{equation}\label{Eq.GWRate}
	\text{GW Event Rate} = \langle V \rangle \langle R \rangle
\end{equation}
where $\langle R \rangle$ is LIGO-Virgo's empirically motivated rate-density estimate~\citep{GW170104,O1BBH,GW150914Rate,GW150914RateSupp,GW170817}, and $\langle V \rangle$ is the average sensitive volume of the epoch's GW search.
We estimate the cosmologically-corrected sensitive volume for each epoch at an SNR threshold of 8, $\langle V_8 \rangle$, using the online distance calculator provided by~\citep{DistanceCalc2017}, so that the average sensitive volume at a given $\rho$ is given by:
\begin{equation}\label{Eq.SensitiveVolume}
	\langle V \rangle = \langle V_8 \rangle \times \left( \frac{\rho}{8} \right)^{-3}
\end{equation}
The three different observing epochs we consider are~\citep{DistanceCalc2013}:  the second Advanced LIGO-Virgo observing run (O2), the third Advanced LIGO-Virgo observing run (O3), and the eventual design sensitivity Advanced LIGO-Virgo observing runs.
For BBH searches, we consider both uniform-in-log and power-law (with a power of -2.35) source-mass distributions~\citep{GW170104,O1BBH,GW150914Rate,GW150914RateSupp} when estimating $\langle V \rangle$ and $\langle R \rangle$.
The differences in the results for these two distributions are negligible, thus we only quote the power-law results in this paper.
For the power-law distribution, we use $\langle R \rangle = 103^{+110}_{-63} \ \text{Gpc}^{-3} \text{yr}^{-1}$~\citep{GW170104}, and averaging over the sensitive volumes of the mass distribution we find $\langle V_8 \rangle_\text{O2} = 0.22 \ \text{Gpc}^3$, $\langle V_8 \rangle_\text{O3} = 0.66 \ \text{Gpc}^3$, and $\langle V_8 \rangle_\text{Design} = 2.3 \ \text{Gpc}^3$.
For BNS, we use the median total mass estimate of $2.8 M_\odot$ for GW170817~\citep{GW170817} when estimating $\langle V \rangle$ and $\langle R \rangle$.
We use $\langle R \rangle = 1540^{+3200}_{-1220} \ \text{Gpc}^{-3} \text{yr}^{-1}$~\citep{GW170817} and $\langle V_8 \rangle_\text{O2} = 0.002 \ \text{Gpc}^3$, $\langle V_8 \rangle_\text{O3} = 0.007 \ \text{Gpc}^3$, and $\langle V_8 \rangle_\text{Design} = 0.03 \ \text{Gpc}^3$.

Combining Eq.~\ref{Eq.Background}, Eq.~\ref{Eq.GWRate}, and Eq.~\ref{Eq.SensitiveVolume}, we compute the expected rate of GW events at each FAR threshold.
In Fig.~\ref{Fig.AllPowerLaws}, we show the fractional increase in GW events expected above each FAR threshold as compared to the baseline FAR threshold of 1 per 100 years.
The most notable result is that increasing the FAR threshold by 5 orders of magnitudes from 1 per 100 years to 1000 per year only increases the number of detectable GW events by approximately a factor of 1.6 for BNS and 1.9 BBH.
In other words, because the GW event rate scales as $\rho^{-3}$, we need to change the $\rho$ threshold by a factor of $\left(\frac{1}{2}\right)^{\frac{1}{3}} \sim 0.8$ to gain a factor of 2 in the number of detectable GW events.
However, changing $\rho$ from $\sim 10$ (corresponding to a FAR threshold of 1 per 100 years) by a factor of $0.8$ increases the FAR contamination by more than 5 orders of magnitude  (see Fig.~\ref{Fig.AllPowerLaws}). 
We emphasize that in practice these numbers only represent the increased number of real GW events as candidates but not necessarily as detections.
To claim any of these additional low-significance GW events as confident detections, they would need to be jointly detected by EM observations.
Thus, the actual increase in the total number of confident GW detections may be lower than this factor of 2 depending on the EM detection efficiency.  

Because the uncertainties in the GW event rate amount to a constant normalization factor (see Eq.~\ref{Eq.GWRate} and Eq.~\ref{Eq.SensitiveVolume}), they do not factor into Fig.~\ref{Fig.AllPowerLaws}.
The only uncertainties that affect this plot are therefore related to the background fit (the following applies to the area/distance/volume localizations discussed in Sec.~\ref{Sec.FalseAlarmCost} as well).
We manually vary the normalization of the total background rate, $\text{FAR}_8$, by up to an order of magnitude, however this only results in negligible uncertainties in the expected rate of detectable GW events.
The effect of the uncertainty regarding the slope, $\alpha$, of our exponential background fits is illustrated by comparing the results for the steeper BNS background to those for the less-steep BBH background.
The relative increase in the rate of detectable GW events is greater for BBH searches than for BNS searches because the range of $\rho$ spanned at these FARs is greater for BBH searches (see Fig.~\ref{Fig.AllPowerLaws}).
Nevertheless, the results of both of these searches are of similar magnitude across all FARs.
Thus, we would only expect the numbers in Fig.~\ref{Fig.AllPowerLaws} to change significantly if the backgrounds for any search were to become drastically more/less heavily-tailed.

\section{Purity of Low-Significance Alerts}\label{Sec.AlertPurity}

\begin{figure}[h!]
    \scalebox{.85}{\includegraphics[width=0.5\textwidth]{\PlotPath/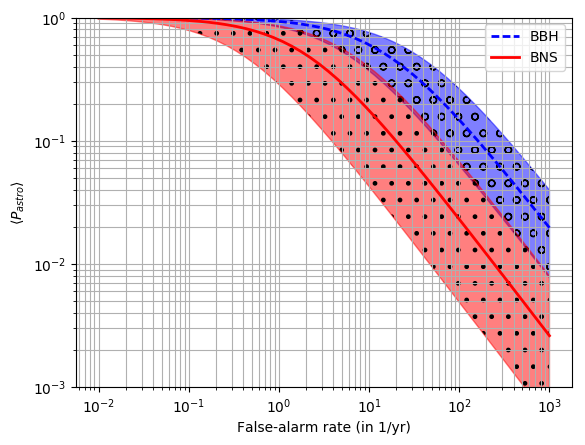}}
    \scalebox{.85}{\includegraphics[width=0.5\textwidth]{\PlotPath/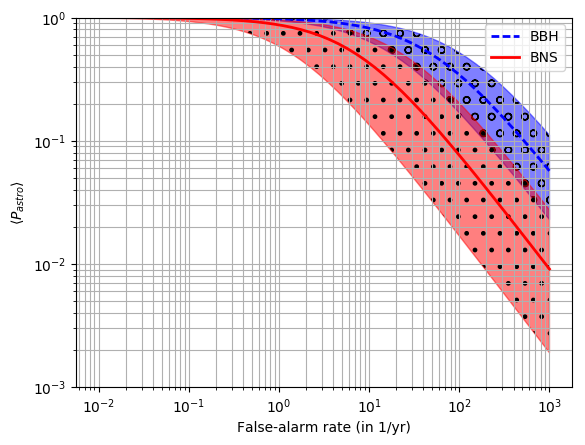}}
    \scalebox{.85}{\includegraphics[width=0.5\textwidth]{\PlotPath/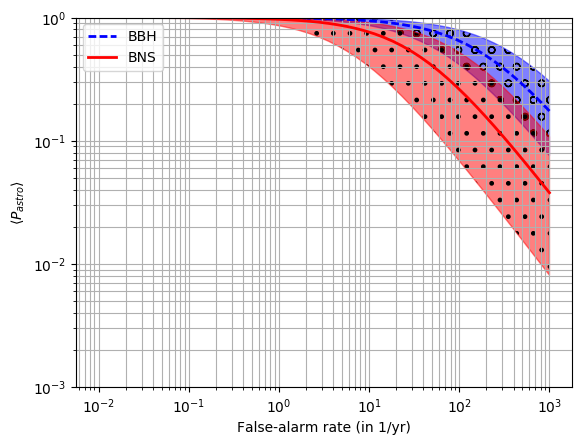}}
    \caption{The average probability of a GW candidate being of astrophysical origin (rather than a false alarm), $\langle P_\text{astro} \rangle$, at each FAR threshold for three observing epochs:  O2 (top), O3 (middle), and design sensitivity (bottom).
    The solid lines correspond to the median published rates $\langle R \rangle$~\citep{GW170104, GW170817} for BBH and BNS events, and the shaded regions correspond to the 90$\%$ credible regions for those same rates.
    Note how the relationship between $\langle P_\text{astro} \rangle$ and FAR changes with each observing epoch, suggesting that FAR is not a consistent measure of purity.
    }
    \label{Fig.ProbSignal}
\end{figure}

\begin{table*}[htb]
    \centering
    \scriptsize
    \caption{The average probability (in \%) of GW alerts being actual GW events ($\langle P_\text{astro} \rangle$), as depicted in Fig.~\ref{Fig.ProbSignal}, at several ad hoc FAR thresholds for various LIGO-Virgo observing epochs.
    We also give the corresponding fractional increase of GW events ($\text{FI}_{GW}$) for these same thresholds. 
    These values correspond to the median published rates for BBH and BNS events~\citep{GW170104, GW170817}.
    Note that the errors on these probabilities corresponding to rate uncertainties can be large (see Fig.~\ref{Fig.ProbSignal}).
    For O2, the probability of a candidate being a GW could degrade by more than an order of magnitude for lower alert thresholds while still not doubling the number of GW events.
    This degradation is less severe as LIGO-Virgo improves and reaches O3 and design levels of sensitivity, suggesting that alert thresholds based on $\langle P_\text{astro} \rangle$ provide more optimal and consistent information for EM follow-up than thresholds based on FAR.
	    }
    \label{Tab.Contamination}
    \begin{tabular}[c]{c||r|r|r|r|r}
        \hline
        Epoch & 1 per 100 years & 1 per year & 1 per month & 1 per week & 1 per day \\
        \hline
        $\langle P_\text{astro} \rangle$ O2 BBH & 99 & 93 & 56 & 24 & 5 \\
        $\langle P_\text{astro} \rangle$ O2 BNS & 99 & 66 & 15 & 4 & 1 \\
        \hline
        $\langle P_\text{astro} \rangle$ O3 BBH & 99 & 98 & 79 & 49 & 13 \\
        $\langle P_\text{astro} \rangle$ O3 BNS & 99 & 87 & 39 & 13 & 2 \\
        \hline
        $\langle P_\text{astro} \rangle$ Design BBH & 99 & 99 & 93 & 77 & 35 \\
        $\langle P_\text{astro} \rangle$ Design BNS & 99 & 97 & 73 & 40 & 9 \\
        \hline
        \hline
        $\text{FI}_{GW}$ BBH & 1.0 & 1.3 & 1.5 & 1.6 & 1.8 \\
        $\text{FI}_{GW}$ BNS & 1.0 & 1.2 & 1.3 & 1.4 & 1.6 \\
    \end{tabular}
\end{table*}

Although it is interesting to explore the fractional increase in GW event rate associated with each FAR threshold, we must take into account that these fractional increases come with absolute costs to observers.
The quantity of interest when issuing alerts to EM astronomers is the fraction of alerts than are actual GW events.
The GW community quantifies this alert purity by calculating the probability that a GW event is of astrophysical origin, $P_\text{astro}$.
$P_\text{astro}$ is measured by comparing the differential rates of GW and background events at a given value of a search statistic~\citep{O1BBH,GW150914Rate,GW150914RateSupp,Farr}.
This quantity has been produced by LIGO-Virgo in offline analyses~\citep{O1BBH,GW150914Rate,GW170104}, however it may be difficult to provide in low-latency because its proper calculation requires a careful measurement of and/or marginalization over the searches' background and GW event rates.
For ease, we approximate the average $P_\text{astro}$ that we'd expect to observe over {\it all events exceeding a given threshold of $\rho$} as
\begin{equation}
	\langle P_\text{astro} \rangle = \frac{\langle V \rangle \langle R \rangle }{\text{FAR} + \langle V \rangle \langle R \rangle}
\end{equation}
where FAR and $\langle V \rangle$ can be given in terms of $\rho$ using Eq.~\ref{Eq.Background} and Eq.~\ref{Eq.SensitiveVolume}, respectively.
In Fig.~\ref{Fig.ProbSignal}, we plot $\langle P_\text{astro} \rangle$ versus the FAR associated with the $\rho$ threshold using the predicted sensitivities for the O2, O3, and design sensitivity observing epochs.
We depict the uncertainties associated with the GW event rate as the shaded regions, while plotting the results for the median published rates as lines.
We give the explicit values of some of these probabilities (corresponding to the median published rates) at several ad hoc FAR thresholds in Table~\ref{Tab.Contamination}.

At low FAR thresholds (like our baseline of 1 per 100 years), we have very high alert purity, meaning there is a great likelihood of success to offset any observational costs incurred to EM observers.
At higher FAR thresholds, the purity is strongly dependent upon the expected GW event rate, i.e., the sensitivity of the detectors.
In O2, LIGO-Virgo observed relatively low GW event rates, meaning high-FAR GW alerts had relatively low probabilities of being real GW events.
However, in more sensitive observing epochs, such as when Advanced LIGO-Virgo reaches design sensitivity, the expected event rate is large enough that even high-FAR GW alerts can have a high purity.
For example, assuming the median event rates~\citep{GW170104, GW170817}, we would have needed a FAR threshold of 2 per year for BNS alerts in O2 to have a $\langle P_\text{astro} \rangle$ of 50$\%$, while at design sensitivity we could instead have a FAR threshold of 30 per year.
Thus, we argue that alert thresholds based on FAR do not convey consistent purity information to EM observers.
The more useful and consistent information needed by astronomers during their cost-reward analysis of follow-up thresholds is a measure of $P_\text{astro}$ that explicitly balances GW event rates against noise rates.
A deeper discussion of the statistical motivation and consequences of thresholding with $P_\text{astro}$ can be found in~\citep{Farr}.
In the absence of explicit estimates of $P_\text{astro}$, Fig.~\ref{Fig.ProbSignal} can be used to map an alert's FAR into estimates of $\langle P_\text{astro} \rangle$.

\section{Observational Implications for EM Follow-up Campaigns}\label{Sec.FalseAlarmCost}

In Sec.~\ref{Sec.AlertPurity}, we describe how raising the FAR threshold decreases the purity of LIGO-Virgo alerts.
However, there is an additional cost associated with low-significance GW events that may directly impact their EM follow-up: events are more poorly localized by the GW detectors as their SNR decreases.
A simple Fisher matrix analysis~\citep{CutlerFlanagan} shows that the uncertainty in GW distance estimates, $\sigma_D$, roughly scales as $\sigma_D \propto \text{SNR}^{-2}$~\citep{FairhurstLocalization}.
More in-depth calculations can be used to show that the angular area uncertainty in GW localization estimates, $\sigma_A$, roughly scales as $\sigma_A \propto \text{SNR}^{-2}$~\citep{SkyAreaScaling}, which agrees with the findings of Monte Carlo studies~\citep{SkyLocRecolored}.
Thus, applying an additional factor of distance squared ($\propto \text{SNR}^{-2}$) to convert the angular area uncertainties to proper area uncertainties, we expect the total uncertainty in GW localization volume, $\sigma_V$, roughly scales as $\sigma_V \propto \text{SNR}^{-6}$, which again agrees with the findings of Monte Carlo studies~\citep{DelPozzoVolume}.

We again assume that the network ranking statistic $\rho \sim \text{SNR}$ for all low-significance GW alerts.
In Fig.~\ref{Fig.AllPowerLaws}, we plot the fractional increase in the GW distance, angular area, and volume localizations for threshold events, as compared to the localization at our baseline FAR of 1 per 100 years.
Similarly to the results for GW event rates (which also scales as a power law in $\rho$), we find that the relative increase in localization is a slowly-varying function of FAR.
Increasing the FAR threshold by 5 orders-of-magnitude from 1 per 100 years to 1000 per year degrades the distance and angular area localizations by less than a factor of 2 and the volume localization by less than a factor 4.
The discussion of the uncertainty in these numbers is identical to the discussion for GW event rate in Sec.~\ref{Sec.SignalRates} above.
Thus, we do not expect low-significance follow-up efforts to experience order-of-magnitude increases in localization costs as compared to those of current threshold events.

We will now discuss how these observational costs will realistically affect EM follow-up.
Here we focus specifically on follow-up procedures, although it should be noted that a serendipitous coincident detection of GW candidates with high-energy telescopes like Fermi/GBM, INTEGRAL/SPI-ACS, Konus/WIND~\citep{GRB170817} may affect the significance of GW candidates~\citep{LindyGRB}.
These instruments have the advantage of continually monitoring a big fraction of the high energy sky, meaning they do not need to be run in follow-up mode. 
Additionally, they are usually subject to backgrounds that are overall quieter than the corresponding ones in optical bands. 
Thus, they present a low-cost means of potentially increasing the significance of GW events in near real time (as was the case with GW170817~\citep{EMFollow170817}).

The EM follow-up observations of GW counterparts are undertaken in stages. 
Transients detected by imaging systems are assessed by spatial location (either 2- or 3-dimensional), broadband spectral characteristics, and light curve temporal evolution.  
These assessments can be accomplished with 2-4 meter aperture telescopes. 
If a viable EM counterpart is detected, large-aperture (8-10 meter class) spectroscopic observations are obtained. 

We consider the impact of a higher false-alarm rate and poorer localization on EM follow-up efforts in three regimes: 
\begin{itemize}
\item{} wide-field surveys, for which follow-up observations amount to re-ordering the sequence in which regions of the sky are observed; 
\item{} galaxy-targeted or other narrow-field imaging programs, which search for transients consistent with GW counterparts; and 
\item{} large-aperture spectroscopic follow-up campaigns, which obtain spectra of individual sources of interest.
\end{itemize}

\subsection{Wide-Field Sky Surveys}

Examples of wide-field imaging systems are the Panoramic Survey Telescope and Rapid Response System (Pan-STARRS)~\citep{MoKa2012}, the Asteroid Terrestrial-impact Last Alert System (ATLAS)~\citep{Ton2011}, the Zwicky Transient Factory (ZTF)~\citep{ZTF}, and eventually, the Large Synoptic Survey Telescope (LSST) \citep{Ivezic2014}.
During the Advanced LIGO-Virgo runs O1 and O2, observations by survey telescopes contributed significantly to the follow-up program for many of the candidates \citep{SmCh2016,SmCh2016b,StTo2017,SmCh2017}, both in estimating the most recent time of non-detection and in observing the fields after a GW alert.

For optical/infrared surveys carrying out high-cadence observations of the entire sky, responding to a GW alert is simply a matter of re-ordering the sequence of observations and perhaps changing broadband filters more rapidly than would otherwise be the case. 
For these systems, localization costs are unimportant, and the primary observational cost is the loss of on-sky efficiency due to the additional filter changes, which require a time overhead that could otherwise be used for observation. 
The acquisition of the images can therefore be accomplished with minimal opportunity cost. 
The scientific opportunity cost of re-ordering the observations should be weighed against several factors, such as the probability of detecting the EM counterparts of low-significance GW alerts and the quality of science that can be extracted from any successful detections.

\subsection{Targeted Imaging Observations}

Targeted imaging observations were used to first detect the optical counterpart to GW170817~\citep{Swope,DECAM,DLT40,LasCumbres,VISTA,MASTER}.
For narrow-field imaging systems that either tile the GW localization region on the sky or target individual galaxies (e.g., ~\citep{GehrelsGalaxy,ArcaviStrategy,DECAM150914}), the EM follow-up observations are often conducted in a target-of-opportunity mode where previously scheduled programs are preempted by the GW follow-up campaign. 
In principle, one could envision dedicating a narrow-field follow-up system entirely to GW follow-up observations.

For narrow-field imaging, the increased alert rate rather than the modest increases in angular localization area (relevant for tiling) or localization volume (relevant for galaxy targeting) dominates the follow-up time requirement.
Thus, we again see that the expected science that can be done with a limited number of additional low-significance GW events will need to counteract the cost of preempted observing programs.

\subsection{Large-Aperture Spectroscopy}

Interrupting the observing program of a large-aperture spectroscopic telescope (such as Keck~\citep{Keck} or Gemini~\citep{Gemini}) to obtain a sequence of spectra for a faint transient is arguably the most costly element in low-significance EM follow-up observations. 
Modest-resolution spectroscopy is extremely valuable for both discrimination and characterization of EM counterparts, but at the same time large aperture telescopes are typically the most over-subscribed resource in the arsenal of follow-up tools.
One final time, we suggest that observers will need to carefully weigh the science output of low-significance GW detections against the cost of studying their EM counterparts.
While there are roughly a factor of 2 more GW events that could optimistically be detected in the EM, it is not immediately clear how the scientific gain of these efforts compare to those of other observing programs sharing the telescopes' resources.

\section{Conclusions}\label{Sec.Conclusions}

The background in gravitational-wave searches for binary systems with LIGO-Virgo falls off exponentially steeply as a function of the detection statistic, while the GW event rate only falls off as a power law. 
As a result, lowering the detection statistic enough to double the number of expected GW detections would inflict an increase of {\it greater than five orders of magnitude} in the false-alarm contamination.
However, the false-alarm rate is not particularly well-suited in determining the scientific merits of follow-up efforts since it contains no information about the expected rate of GW events during the observing epoch.
Instead, a quantity that compares the relative rates of GW events and false-alarms, such as the probability of an alert being astrophysical ($P_\text{astro}$), is the more appropriate metric for setting consistent follow-up thresholds.
If LIGO-Virgo were to provide estimates of $P_\text{astro}$ for alerts in real-time, the observing facilities could then threshold on it directly.
In lieu of such information, estimates of $P_\text{astro}$, such as $\langle P_\text{astro} \rangle$ described in this paper, could be used as a proxy.

Optimistically, the resources are available for the joint GW-EM community to roughly double the number of GW detections by following-up low-significance GW candidates with EM observations.
Nevertheless, the trade-off between the scientific value of these additional detections and the effort required to obtain them is not immediately or universally clear.
The joint GW-EM community should always attempt to maximize the scientific output of its observational efforts.
In the age of multi-messenger astronomy, it is now in the enviable position of having to quantify the point of diminishing returns for additional low-significance detections.

\section{Acknowledgments}

The authors acknowledge the support of the National Science Foundation and the LIGO Laboratory.
LIGO was constructed by the California Institute of Technology and Massachusetts Institute of Technology with funding from the National Science Foundation and operates under cooperative agreement PHY-0757058.
MC is supported by the David and Ellen Lee Postdoctoral Fellowship at the California Institute of Technology.
We would also like to thank Hsin-Yu Chen, Kwan Yeung Ng, Yiwen Huang, Robert Eisenstein, Satya Mohapatra, Steve Drasco, Peter Shawhan, Alex Nitz, Tito Dal Canton, Christopher Berry, Thomas Dent, Reed Essick, and the LIGO-Virgo EM follow-up working group for useful comments and discussion. CWS is grateful to Harvard University and to the US Department of Energy (through grant DE-SC000788) for their support of this effort. 
This is LIGO document number P1800042.


\bibliography{refs}

\end{document}